\documentclass[11pt]{article}

\usepackage[utf8]{inputenc}
\usepackage{lipsum}
\usepackage[space]{grffile}
\usepackage{graphicx}
\usepackage{url}
\usepackage{amsmath}
\usepackage{amssymb}
\usepackage{upgreek}
\usepackage{multirow}
\usepackage{verbatim}

\usepackage{mathtools}

\usepackage{tikz}
\usepackage{pgfplots}
\usepackage{tabularx}
\usepackage{esvect}
\usepackage{array, boldline, makecell, booktabs}\newcolumntype{\$}{>{}}

\usepackage{float}  
\usepackage{tikz,tikz-3dplot}
\usetikzlibrary{calc,matrix,fit}

\usetikzlibrary{positioning,arrows.meta,quotes}
\usetikzlibrary{shapes,snakes}
\usetikzlibrary{bayesnet}
\tikzset{>=latex}

\definecolor{barBlack}{HTML}{000000}
\definecolor{barGrey}{HTML}{808080}

\newcommand{\R}{\mathbb{R}}

\usepackage[ruled,vlined]{algorithm2e}
\SetCommentSty{mycommfont}
\usepackage{algpseudocode}

\newcommand{\known}[1]{\mbox{$\sigma(#1)$}}
\newcommand{\unknown}[1]{\mbox{$\overline{\sigma}(#1)$}}

\SetKwInput{KwInput}{Input}                % Set the Input
\SetKwInput{KwOutput}{Output}              % set the Output

\begin{document}

\title{\bf Rank-Preference Consistency as the Appropriate Metric for Recommender Systems}
\author{{\bf Tung D.\ Nguyen} and  {\bf Jeffrey Uhlmann}\vspace{4pt} \\
Dept.\ of Electrical Engineering and Computer Science\\
University of Missouri - Columbia}
\date{}
\maketitle

\begin{abstract}
ABSTRACT: In this paper we  argue that conventional unitary-invariant measures of recommender system (RS) performance based on measuring differences between predicted ratings and actual user ratings fail to assess fundamental RS properties. More specifically, posing the optimization problem as one of predicting {\em exact} user ratings provides only an indirect suboptimal approximation for what RS applications typically need, which is an ability to accurately predict user {\em preferences}. We argue that scalar measures such as RMSE and MAE with respect to differences between actual and predicted ratings are only proxies for measuring RS ability to accurately estimate user preferences. We propose what we consider to be a measure that is more fundamentally appropriate for assessing RS performance, {\em rank-preference consistency}, which simply counts the number of prediction pairs that are inconsistent with the user's expressed product preferences. For example, if an RS predicts the user will prefer product A over product B, but the user's withheld ratings indicate s/he prefers product B over A, then rank-preference consistency has been violated. Our test results conclusively demonstrate that methods tailored to optimize arbitrary measures such as RMSE are {\em not} generally effective at accurately predicting user preferences. Thus, we conclude that conventional methods used for assessing RS performance are arbitrary and misleading. 
\end{abstract}

\section{Introduction}

In this paper we argue that measures of performance focused on predicting user ratings should be viewed as suboptimal proxies for a deeper and more fundamental metric relating to users' {\em rank preferences}. Specifically, if a recommender system predicts that a user will give a rating of X for product A, and a rating of Y for product B, then if X>Y we should expect the user to prefer product A over product B. It is difficult to imagine any sensible argument to the contrary. Therefore, the fundamentally ``natural'' measure for assessing the quality of a given RS should measure the extent to which its predictions are consistent with the expressed rank preferences of users. 

Heuristically, a measure of how accurately an RS predicts the {\em exact} ratings a user has given for a set of withheld products should tend to correlate with the consistency of those predictions with the user's actual rank-order preferences among those products. However, scalar measures of differences between predicted ratings and actual ratings, e.g., RMSE and MAE, are only meaningful to the extent that they actually {\em do} estimate rank-order preferences of the user. For example, when deciding to recommend a particular product to a user Alice, is the goal to accurately predict that she will give it a particular high rating? Or is it to accurately predict the extent to which she will prefer it over other products she has rated? These questions are obviously related, but we argue that the latter is more directly relevant than the former. 

The outline of the paper is as follows: We begin with background on consistency-based RS methods, which provide provable properties relating to rank-order estimation. We then provide background on SVD-based methods and newer AI-based approaches. We then empirically compare these methods to assess their relative effectiveness at generating predictions that are consistent with rank-order user preferences of withheld products. This is measured in the simplest and most direct way possible: {\em How many pairs of RS predictions incorrectly indicate a user will prefer product X over Y when the user actually rated Y over X?}

\section{Consistency-Based RS Methods}

In this section we provide background on the unit-consistent (UC) and shift-consistent (SC) recommender system frameworks, both of which provide provable performance guarantees in the form of the {\em consensus-order property} \cite{acmrs}. This property simply says that if all users prefer product A over product B, then the RS can never recommend product B over product A to a new user \cite{acmrs}. This consensus-order property would seem to be a minimum performance expectation of any recommender system, but methods that are tailored to optimize generic performance measures such as RMSE cannot guarantee this property. It is argued in \cite{acmrs} that this consensus-order property is so fundamental as to represent an {\em admissibility criterion} for any proposed recommender system. 

Currently there are two formally proven admissible RS approaches, each of which guarantees the consensus order property (and more) by making recommendations based on a single consistency constraint. The first enforces {\em unit consistency} \cite{acmrs}. This property can be interpreted as saying that if the units (e.g., centimeters versus meters) on input and output variables are changed, i.e., scale factors are applied to the rows and columns of the rating matrix/table, then the new solution will be identical to the previous solution except given in the new units. For illustration, a unit-consistent (UC) recommender system implicitly assumes that if a user Alice produces product ratings that are consistently 10\% higher than those from Bob for the same products, then if Bob were to give a new product a rating of 70/100, it is reasonable to predict that Alice is likely to give it a rating that is 10\% higher than Bob's, i.e., 77/100. 

The second provably admissible approach we consider is based on {\em shift consistency} \cite{recsys-sc}. This can be interpreted as saying that if all entries in any given row or column of the rating matrix are shifted by a constant value, then the new solution will be identical to the previous solution but retaining the applied shifts. Thus, a shift-consistent (SC) recommender system implicitly assumes that if Alice consistently gives ratings that are 1 unit higher than those of Bob, then if Bob gives a new product a rating of 7/10, it is reasonable to expect Alice to give that product a rating of 8/10. 

The UC and SC constraints are independently sufficient to guarantee the consensus-order property and thus provide complete RS solutions. The fact that a single constraint completely determines all recommendation decisions has important implications: 
\begin{itemize}
    \item Unlike virtually all other approaches described in the literature, consistency-based methods do not require any hyperparameters. This is because they are {\em complete} in the sense that their performance is uniform across datasets, i.e., they do not rely on statistical modeling. SVD-based methods, however, rely on a variety of heuristic methods for choosing the values of critical hyperparameters. These heuristics can be very sensitive to small-sample errors because they rely on statistics determined from the dataset on which they will operate. Most AI-based methods are similarly opaque in the sense that generalization properties are data dependent rather than intrinsic. Consistency-based methods, by contrast, can be fully understood from the rigorous properties guaranteed by its single governing consistency constraint. 
    \item The fact that consistency-based methods have no heuristic or data-specific dependencies imbues them with a kind of intrinsic  ``fairness'' in the sense that their recommendations are determined entirely by a single well-understood consistency constraint. By contrast, alternative methods are explicitly open to potential manipulation via their choice of hyperparameter values and by a variety of subtle under-the-hood design choices -- {\em some of which may admit intentional backdoor opportunities for surreptitious manipulation}. Because UC and SC are relatively very simple matrix completion algorithms with no hyperparameters, implementations can be readily analyzed and fully understood. 
\end{itemize}
From a practical perspective, the most important property of the UC/SC framework is that its benefits derive from an algorithm that is optimally efficient in the sense that its preprocessing time is linear in the number of filled entries in the rating table (i.e., is {\em sparse-optimal}), and its completion time for a missing element is $O(1)$. It also generalizes elegantly to higher dimensions, e.g., allowing each product and/or user to be represented as a vector of attributes. 
A full exposition of the consistency-based RS framework can be found in \cite{acmrs}.

\section{SVD-Based RS Methods}

The conventional mathematical formulation of the matrix-completion problem as applied to recommender systems is to find the minimum-rank completion of the user-product ratings table/matrix using a generalized singular value decomposition (SVD) \cite{firstSVD}. As with most other RS approaches, many variants have been examined that are distinguished by their choice of side constraints and hyperparameters. Examples include \cite{szl,hardt,cv-background1,cv-object,cv-visual,tnnr1, tnnr2,cv-structure,low-tt-rankness}, with RS applications including music recommendation \cite{music-SVD}, collaborative information retrieval \cite{hybridsvd}, and social recommendation \cite{socialRS}.

The distinguishing mathematical property of the SVD is that it obtains a unitary-invariant set of nonnegative singular values. In other words, use of the SVD implicitly assumes that the salient properties of a system of interest are preserved under unitary/orthogonal ``rotation'' transformations. In the context of recommender systems, this implies that information contained in a vector of user ratings is meaningfully preserved if transformed by an arbitrary orthogonal matrix. It similarly implies that information contained in a vector of ratings given by different users for a particular product is meaningfully preserved if transformed by an arbitrary orthogonal matrix. 
It is argued in \cite{acmrs} that the ``mixing'' of user/product ratings under rotation-type transformations is not meaningful within the RS problem.  

Whether or not unitary/orthogonal invariance is meaningful with respect to the rows and columns of of a table of user/product ratings, the preservation of that invariance via use of the SVD will tend to optimize all unitary-invariant norms, e.g., RMSE. Thus, if it is conjectured that unitary invariance is a fundamental property of the RS problem, then we should expect use of the SVD to also optimize the capability of a system to accurately estimate/distinguish user rank preferences, which we argue {\em is} fundamental to the RS problem. Thus, our test results not only provide information relevant to the efficacy of SVD-based RS methods, but also more generally to the relevance of unitary-invariant measures of performance, such as RMSE, for evaluating RS methods. To this end, we examine the Ranking-SVD method \cite{ranking-SVD}, which is specifically claimed to provide more accurate rank-reference predictions than other SVD-based approaches.

\section{AI Approaches to RS: GLocalK}

Most current AI-based RS approaches use deep learning methods to discover latent features of the underlying user-product distribution. Examples include GAN-based \cite{AI-RS-1}, Transformer-based RS \cite{AI-RS-3}, and GNN-based \cite{AI-RS-2, AI-RS-4}. Among these, GLocalK \cite{GLocalK} has been shown to perform best according to RMSE on two of the standard benchmark datasets (MovieLens and Douban-Monti). Although there is no reason to believe it or any other AI-based method is strictly admissible (i.e., that the underlying RS model can never recommend product A over product B to a new user if all existing users rate B over A), that doesn't necessarily imply it cannot approach near-optimal performance in many or most practical settings. 

As discussed in \cite{acmrs}, the recommender system problem superficially seems to fall squarely in the domain of artificial intelligence. This is because the RS problem seems to require a model of human psychology that can be instantiated based on each user's ratings of a set of products. However, the authors argue that the intuitively compelling admissibility criterion is nearly sufficient to yield a complete and unique solution to the RS problem. This would seem to imply that the capability of AI-based models to predict rank preferences will be suboptimal up to the limit at which it approaches admissibility. If this is true, then it could be conjectured that the performance of UC/TC may define the optimal limit for {\em any} AI-based system. This question motivates our consideration of GLocalK for our tests.

\section{Experimental Results}

In this section we provide test results using rank-preference consistency as the measure of performance to compare six RS methods. We use the MovieLens-1M dataset \cite{Movielens100k+1M} and the Douban-Monti Dataset \cite{monti}, both of which contain user-product ratings given on a scale of 1 to 5. The statistics for these datasets are as follow:

\begin{table}[ht]
\centering
\caption{Statistics Summary of MovieLens 1M and Douban Datasets}
\label{tab:dataset-summary}
\begin{tabular}{@{}lcc@{}}
\toprule
Statistic          & MovieLens 1M     & Douban \\ \midrule
Number of Users    & 6,040            & 3,000 \\
Number of Products & 3,706            & 3,000 \\
Sparsity           & 95.53\%          & 98.49\% \\
Number of Entries  & 1,000,209        & 136,000 \\ \bottomrule
\end{tabular}
\end{table}

To assess relative discrimination power, we select a set of K users, and for each user we withhold a pair of products with specific distinct ratings. The methods tested include the following:
\begin{enumerate}
   \item Unit Consistent (UC) - This method provably satisfies the consensus-order property, which should produce predictions that preserve rank-order consistency.
   \item Shift Consistent (SC) - This method is closely related to UC, and it also satisfies the consensus-order property. We expect the UC and SC methods to perform similarly (and conjecture to be optimal) according to the rank-preference consistency metric.
   \item GLocalK - This is a representative example of an AI RS approach. GLocalK first pre-trains an auto encoder to transform the data space to capture the feature space for item interations, then it fine-tunes the pre-trained auto encoder with the original rating matrix to capture the user-item characteristics \cite{GLocalK}. It is computationally more expensive to preprocess (train) than UC and SC, and does not provide any provable performance properties (e.g., consensus order), but it potentially exploits a larger set of data features that may translate to superior practical performance.
   \item SVD (1, 2, and 3) - SVD is the most widely used mathematical basis for RS prediction. \cite{ranking-SVD} introduces a SVD method that is tailored to more accurately provide the top-N user rank-preferred products for a given value of N than conventional SVD-based methods. We consider three variants of this method, each of which retains a specific fraction of the set of singular values. We should expect the use of more singular values to yield better performance than fewer, but at a significantly higher computational cost.
\end{enumerate}

We begin with product pairs for which users have rated one product as 1 and the other as 5. These pairs should be the easiest for an RS to distinguish with respect to user preference, i.e., provide predicted ratings that preserve each user's rank-order preference. More specifically, if a user gives the highest rating of 5 to product A, and the lowest rating of 1 to product B, then a predicted rating for A that is lower than that of B would represent a significant failure. Our metric simply counts the number of discordant prediction pairs for each method. 

As can be seen in Figure 1, UC and SC perform comparably, as expected, and produce the smallest numbers of discordant pairs. However, GLocalK performs comparably, thus hinting that all three methods may exhibit near-optimal performance. The relative performances of the three SVD variants are consistent with the expectation that retaining more singular values will yield fewer discordant pairs, though all three variants perform significantly worse than UC, SC, and GLocalK.

\begin{figure}[ht]
  \centering
  \begin{minipage}{.5\textwidth}
    \centering
    \includegraphics[width=\linewidth]{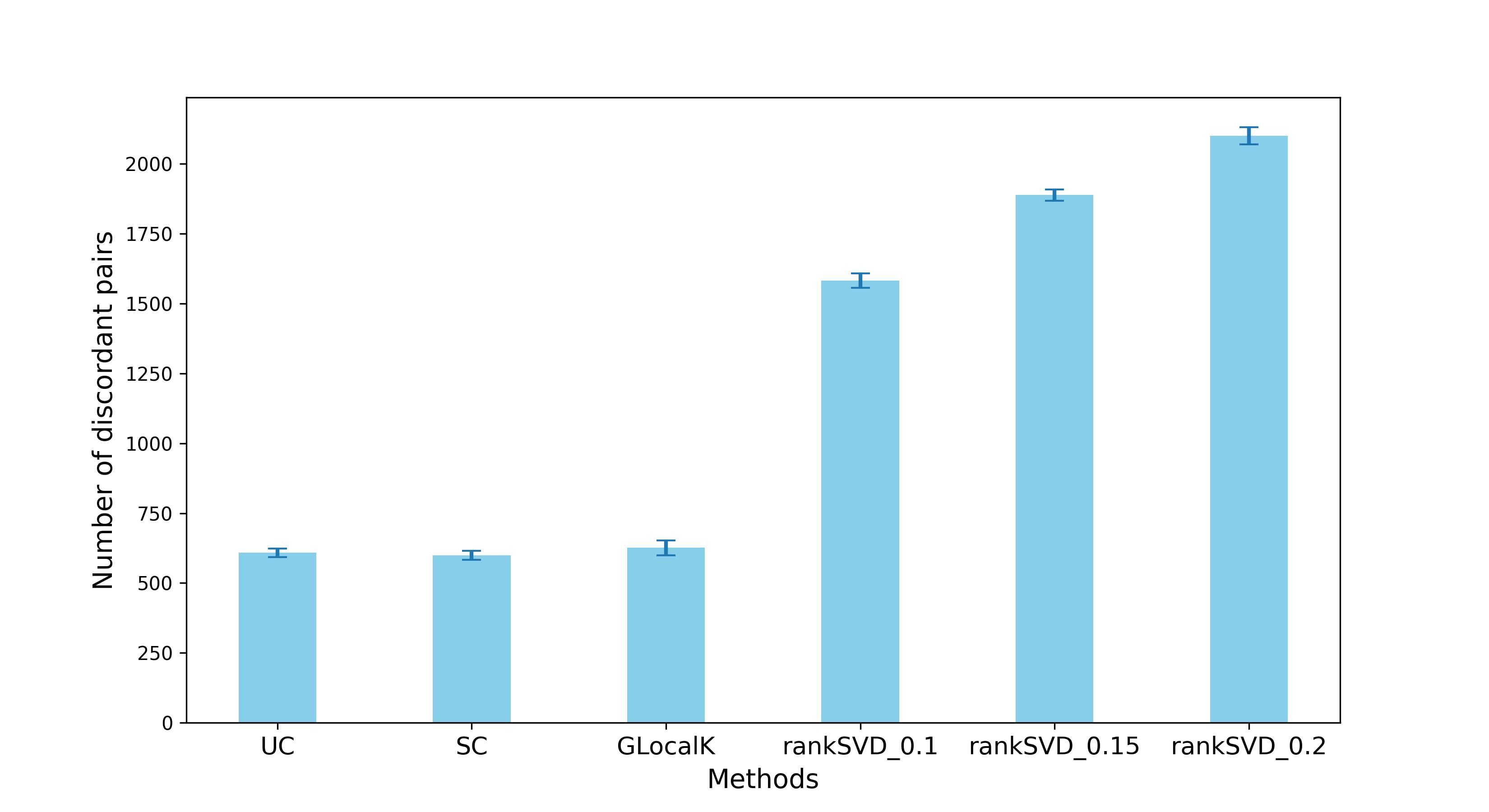}
  \end{minipage}\hfill
  \begin{minipage}{.5\textwidth}
    \centering
    \includegraphics[width=\linewidth]{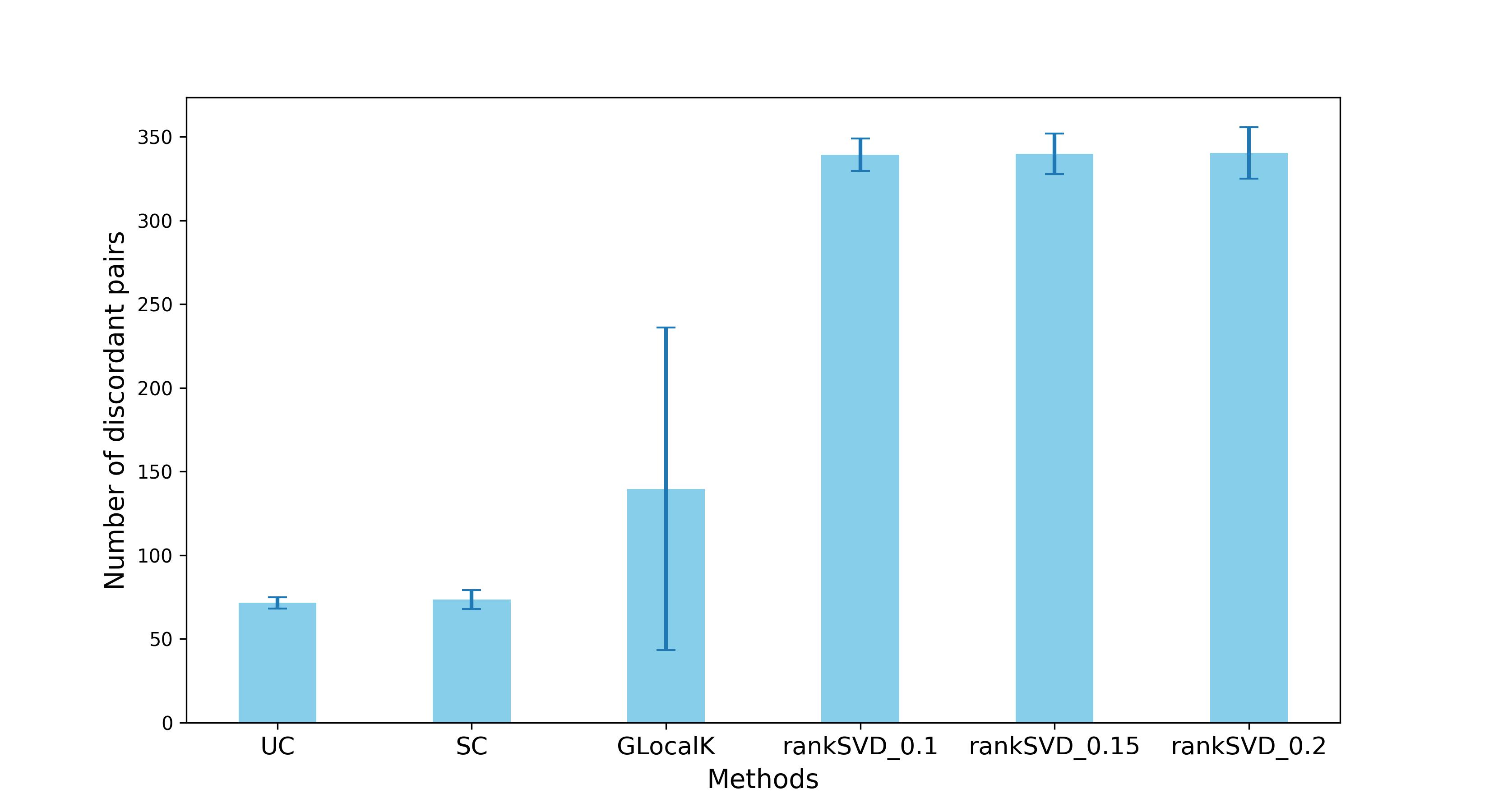}
  \end{minipage}
  \caption{Distinguishing rating 5 from rating 1. (Left) ML-1M dataset, (Right) Douban dataset}
\end{figure}

Figure 2 gives results for distinguishing rank-order preference between product pairs in which one is rated 5 and the other 2. In other words, the difference in ratings is smaller than that of the first test, i.e., less easy to distinguish, so we should expect all methods to produce more discordant pairs. Figure 2 confirms this suspicion, but it also shows qualitatively similar relative performance among the methods.

\begin{figure}[ht]
  \centering
  \begin{minipage}{.5\textwidth}
    \centering
    \includegraphics[width=\linewidth]{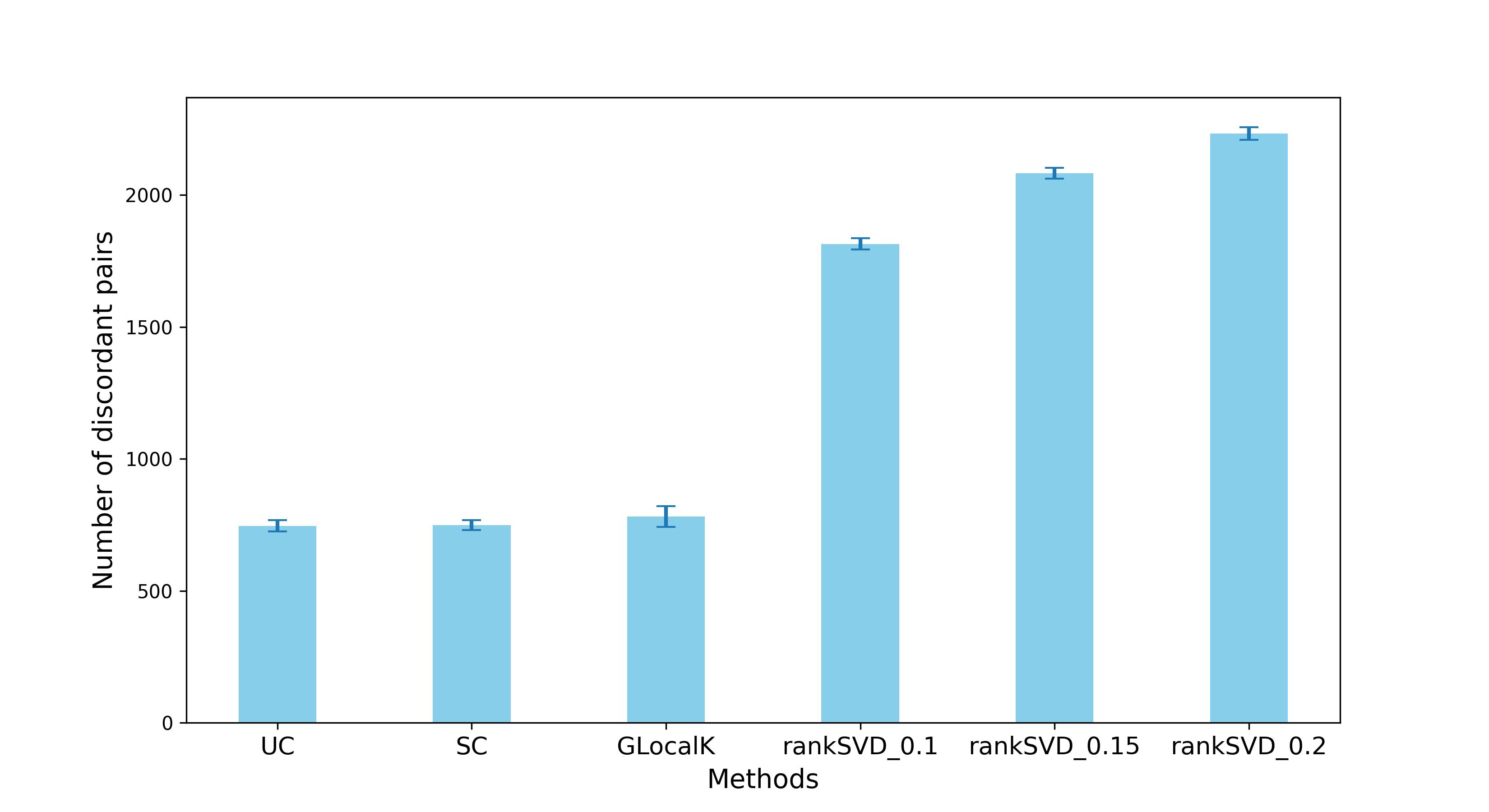}
  \end{minipage}\hfill
  \begin{minipage}{.5\textwidth}
    \centering
    \includegraphics[width=\linewidth]{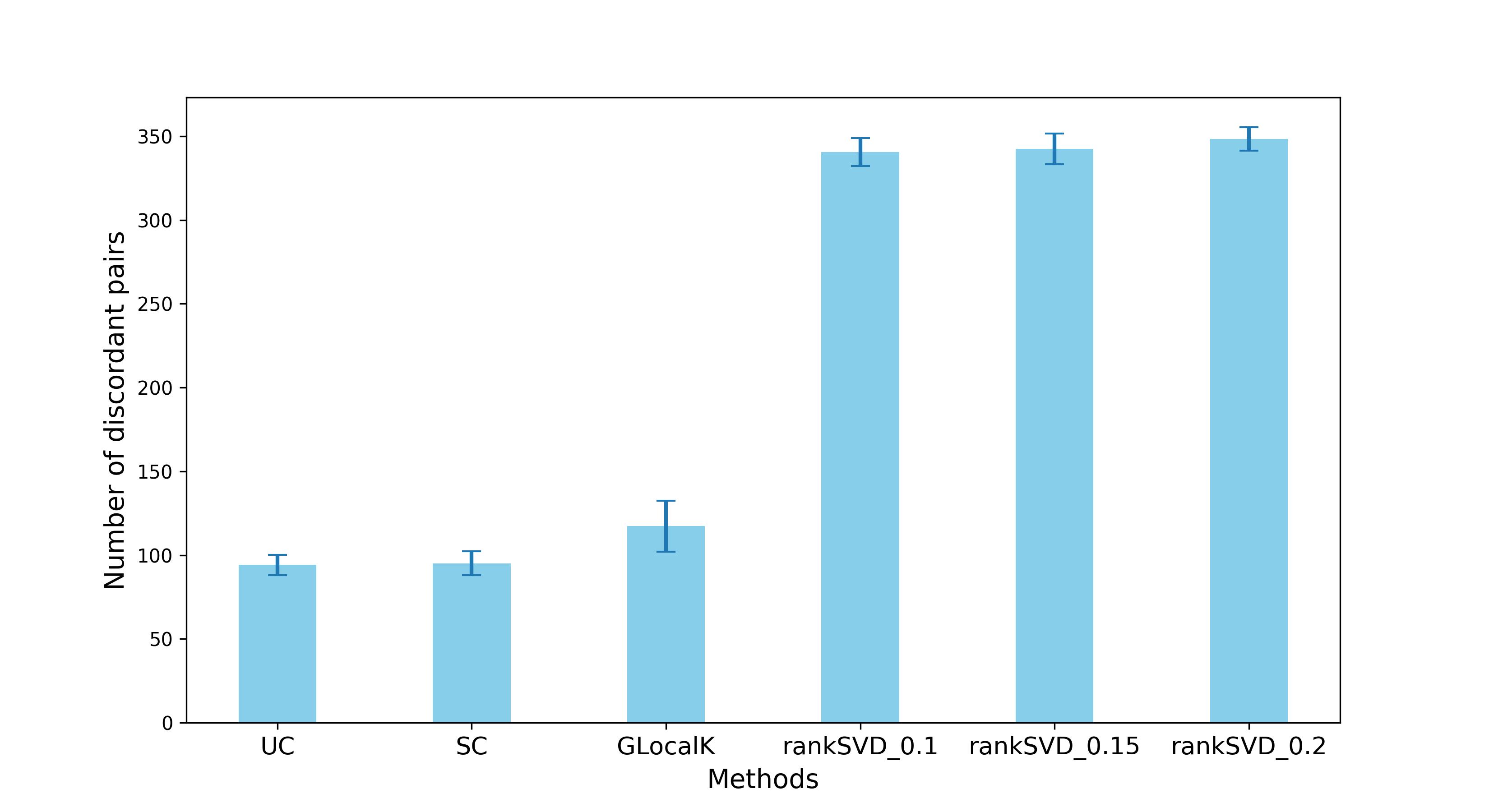}
  \end{minipage}
  \caption{Distinguishing rating 5 from rating 2. (Left) ML-1M dataset, (Right) Douban dataset}
\end{figure}

Figure 3 gives results for distinguishing rank-order preferences between product pairs in which one is rated 5 and the other 3, further increasing the challenge of preserving rank-preference consistency. As before, we should expect all methods to produce more discordant pairs, and this is confirmed. The relative performances of the different methods remain qualitatively similar to the previous tests, though the advantage of UC, SC, and GLocalK over the SVD variants is decreasing. This should not be surprising because all methods should perform equivalently in the non-informative limit of products for which users cannot distinguish a preference within the resolution/granularity of the available rating values. 

\begin{figure}[ht]
  \centering
  \begin{minipage}{.5\textwidth}
    \centering
    \includegraphics[width=\linewidth]{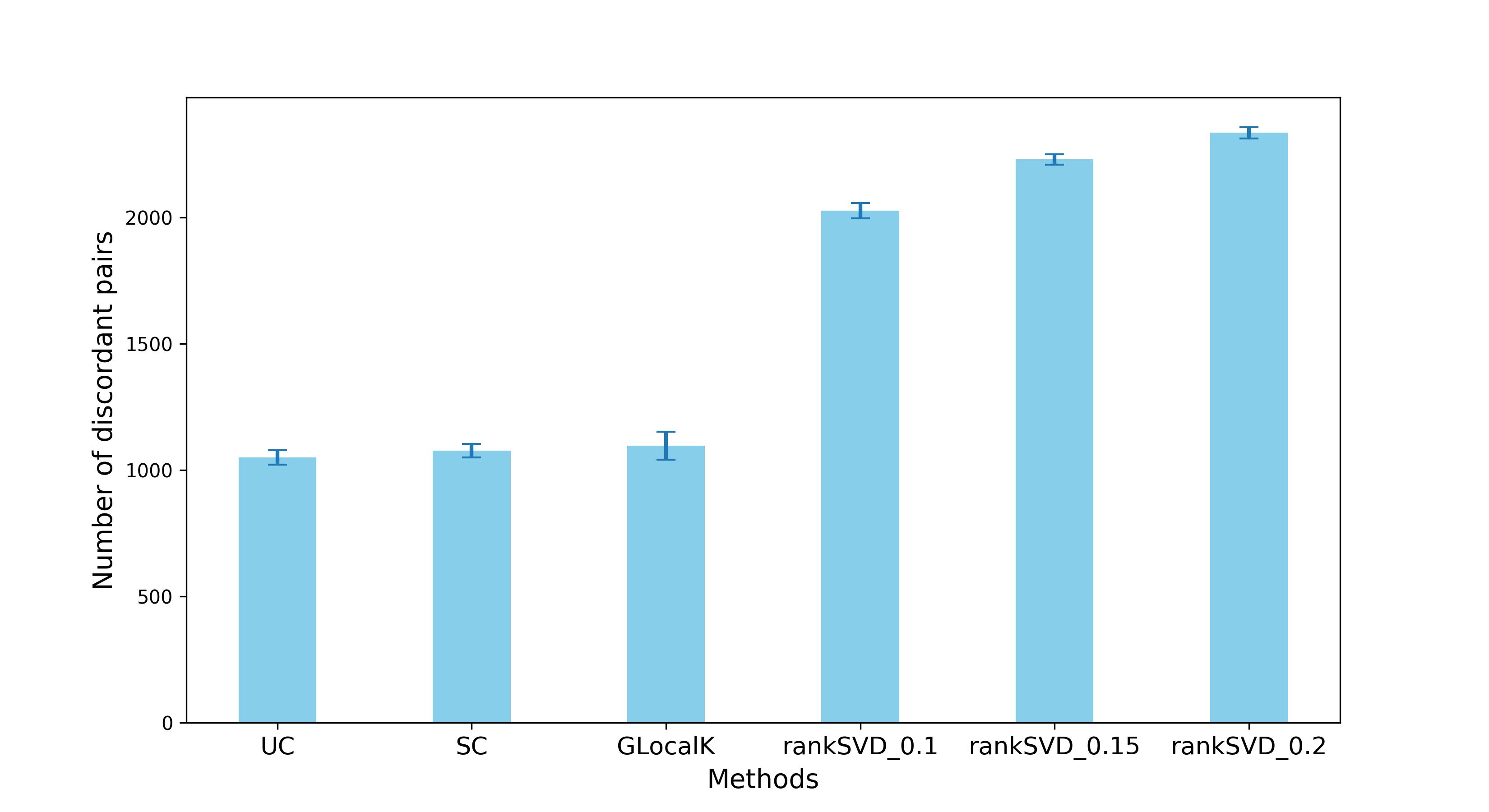}
  \end{minipage}\hfill
  \begin{minipage}{.5\textwidth}
    \centering
    \includegraphics[width=\linewidth]{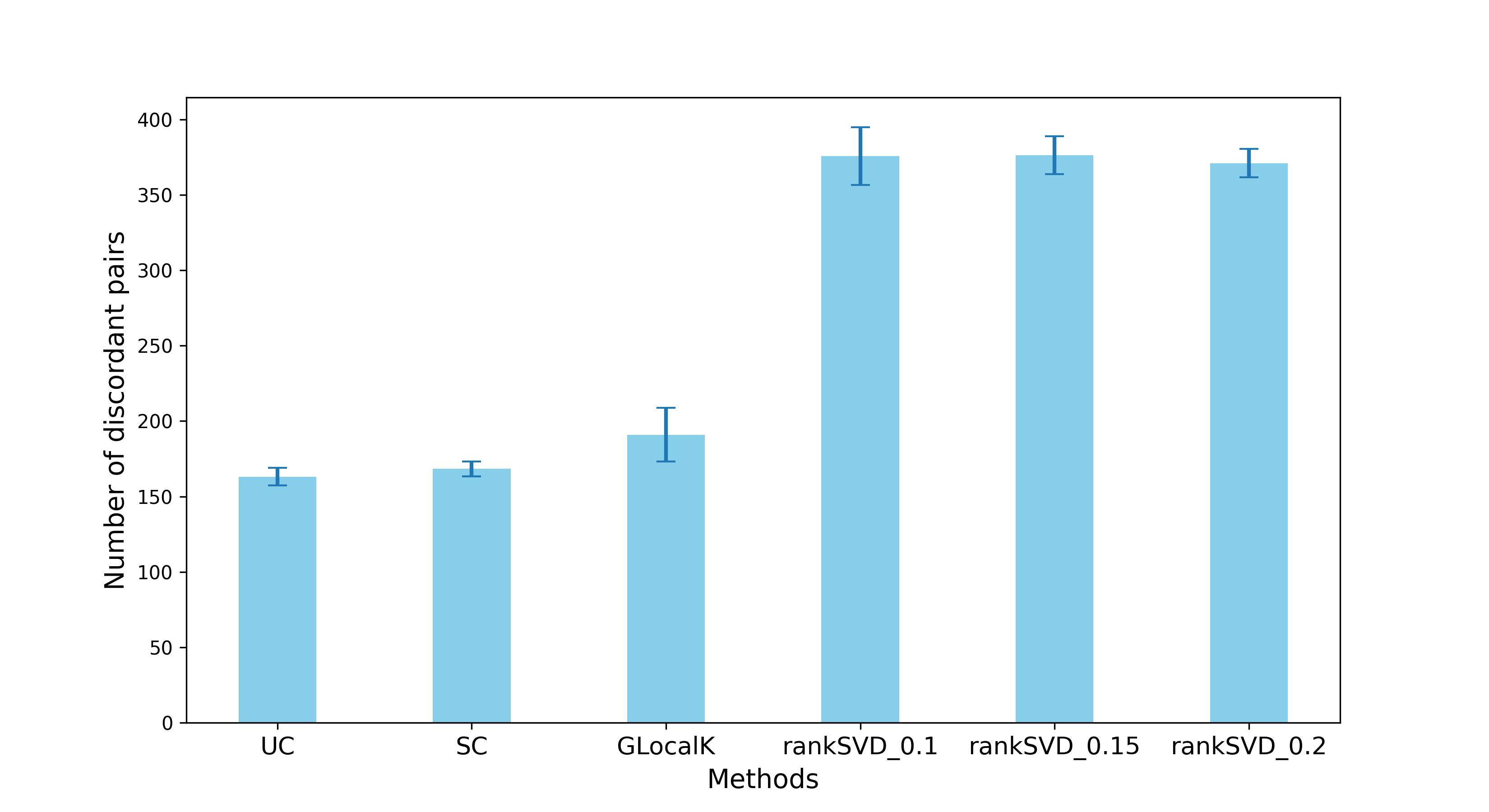}
  \end{minipage}
  \caption{Distinguishing rating 5 from rating 3. (Left) ML-1M dataset, (Right) Douban dataset}
\end{figure}

Figure 4 represents the limit of minimal distinguishability in which the withheld pairs have ratings that differ by 5-4=1. The trend seen from the previous three tests continues here: UC, SC, and GLocalK perform comparably but produce significantly fewer discordant pairs than all of the SVD variants. In summary, all methods produce numbers of discordant pairs that increase inversely with respect to the difference in ratings of withheld pairs, but the relative rank performance of the methods remains stable. 

\begin{figure}[ht]
  \centering
  \begin{minipage}{.5\textwidth}
    \centering
    \includegraphics[width=\linewidth]{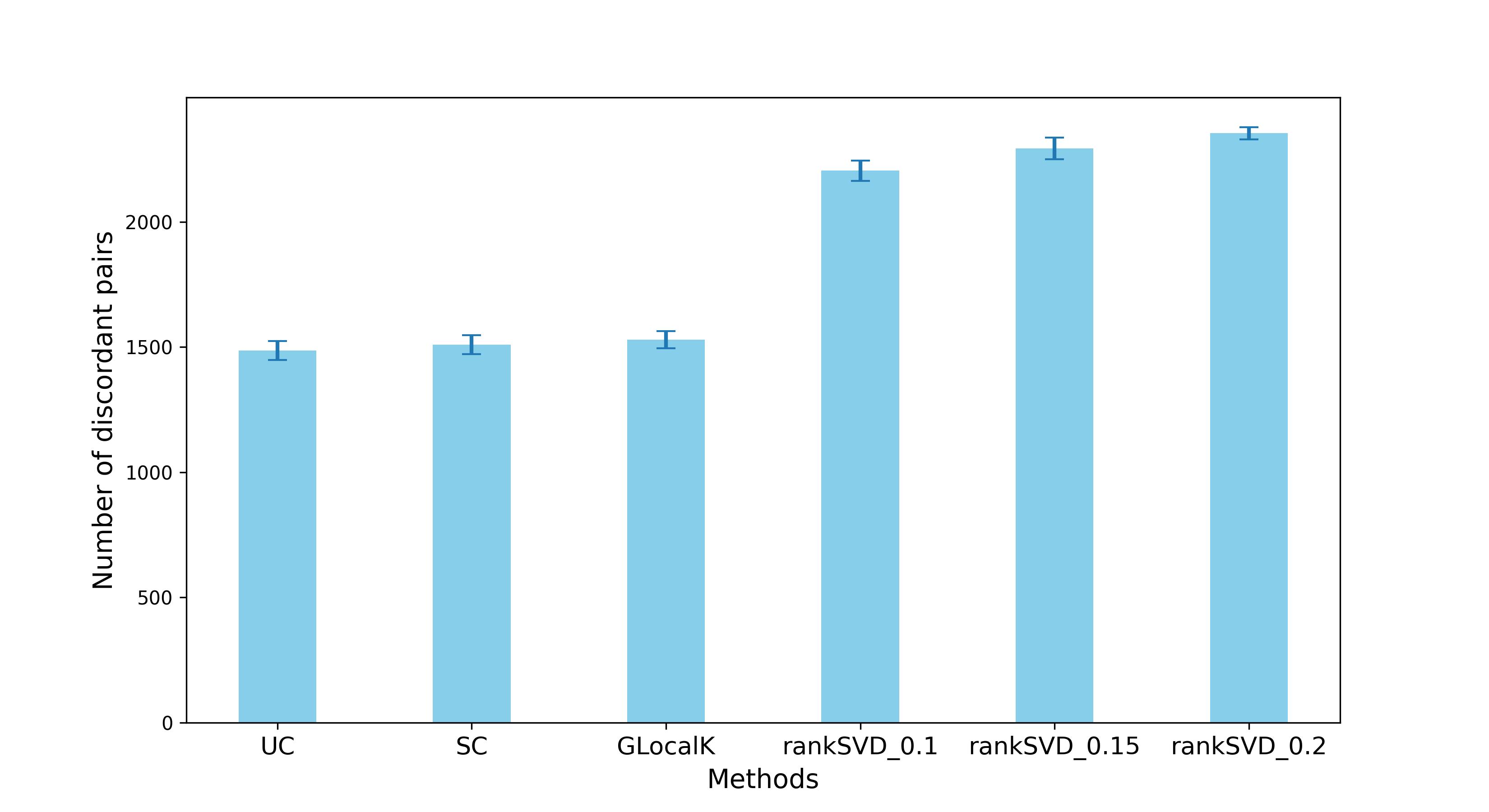}
  \end{minipage}\hfill
  \begin{minipage}{.5\textwidth}
    \centering
    \includegraphics[width=\linewidth]{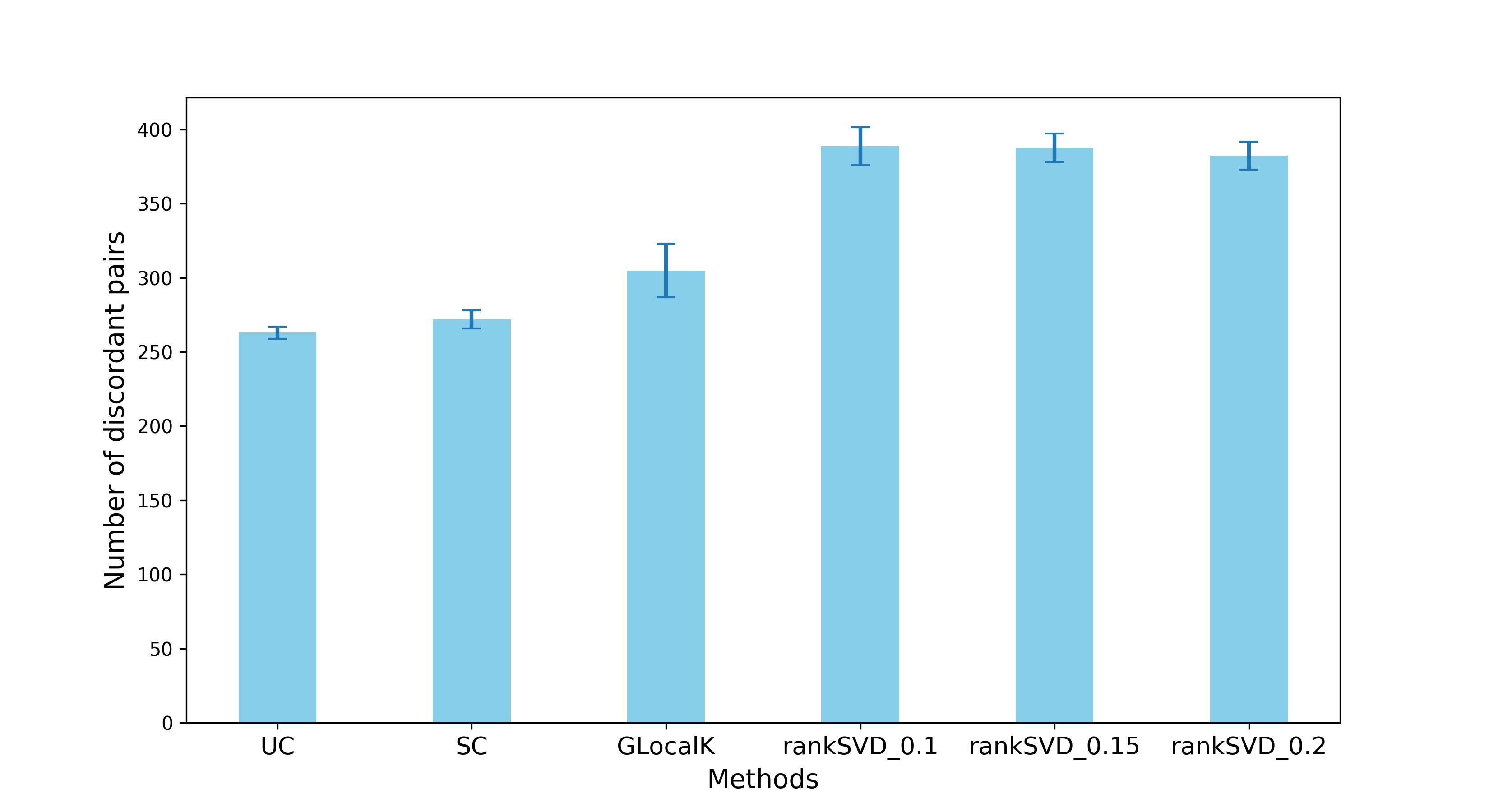}
  \end{minipage}
  \caption{Distinguishing rating 5 from rating 4. (Left) ML-1M dataset, (Right) Douban dataset}
\end{figure}

To summarize, UC, SC, and GLocalK perform comparably across the four tests, despite the fact that UC and SC offer fast $O(n)$ preprocessing computational complexity for a dataset with $n$ ratings, whereas GLocalK requires substantially more computation time for its preprocessing / training phase. The SVD variants require preprocessing that is $O(n k^2)$, where $k$ is the number of retained singular values. Even for large $k$, however, our results show that SVD provides relatively poor rank-preference consistency. This is illuminating because SVD-based methods are typically competitive with or outperform other methods according to RMSE and other generic unitary-invariant performance measures. 

\section{Conclusion}

We have proposed that counting the number of preference-discordant pairs, which is equivalent to the Kendall-Tau metric \cite{kt-metric} when normalized, is a more natural and fundamental measure of RS performance. Our argument for this choice of measure is based on the premise that the objective a recommender system is to provide relative-preference information. More specifically, we argue that it is more important for a given RS to correctly predict on average whether a given user will prefer a given product over another given product than to minimize the mean squared difference (or other arbitary measure) between ratings and predictions.

As we have discussed, a method optimized to accurately predict user ratings should also be expected to give predictions that are consistent with users' rank-order preferences. Our test results show that SVD-based methods, which tend to optimize unitarily invariant scalar measures such as RMSE, do not strongly provide rank-preference consistency. This is consistent with our conjecture that unitary invariance is not a property that is fundamental to the recommender system problem. More generally, this implies that unitary invariant measures of performance, e.g., RMSE and MAE, are not suitable measures of performance for recommender systems.

\section{Acknowledgement}
We thank Quynh Nguyen (Suffolk University) for constructive feedback and assistance in the conducting of our experiments.

\end{document}